\documentclass[pra,aps,superscriptaddress,twocolumn,floatfix,10pt]{revtex4-2}

\usepackage{ragged2e}

\usepackage{dsfont}

\usepackage{mathtools}
\usepackage{amsmath}
\usepackage{amssymb,mathrsfs}
\usepackage{graphicx}
\usepackage{epstopdf}
\usepackage{mathtext}
\usepackage{yfonts}
\usepackage{bm}

\usepackage{soul}

\usepackage{color}
\usepackage[colorlinks=true,linkcolor=blue]{hyperref}

\begin{document}

\title{Exceptional light propagation via generalized bulk-edge correspondence}

\author{Heitor~da~Silva}
\email{heitor.silva@uv.es}
\affiliation{Instituto de Ciencia de los Materiales, Universidad de Valencia, Catedr\'atico J. Beltr\'an, 2, Paterna, 46980, Spain}

\author{Sergey~K.~Ivanov}
\affiliation{Instituto de Ciencia de los Materiales, Universidad de Valencia, Catedr\'atico J. Beltr\'an, 2, Paterna, 46980, Spain}

\author{Isaac~Su\'{a}rez}
\affiliation{Instituto de Ciencia de los Materiales, Universidad de Valencia, Catedr\'atico J. Beltr\'an, 2, Paterna, 46980, Spain}

\author{Jos\'{e}~R.~Salgueiro}
\affiliation{Instituto de F\'{i}sica e Ciencias Aeroespaciais (IFCAE), Universidade de Vigo, Campus de As Lagoas, E-32004 Ourense, Spain}

\author{Albert~Ferrando}
\email{albert.ferrando@uv.es}
\affiliation{Instituto de Ciencia de los Materiales, Universidad de Valencia, Catedr\'atico J. Beltr\'an, 2, Paterna, 46980, Spain}

\begin{abstract}
In topological photonics, the bulk--edge correspondence is conventionally imported by direct analogy with electronic systems, overlooking the fundamentally distinct spacetime symmetries of Maxwell's and Schr\"{o}dinger's equations. In this work, we challenge this prevailing paradigm by demonstrating that non-trivial bulk topology alone is insufficient to guarantee localized edge states in photonic platforms. Using a Su–Schrieffer-Heeger-inspired photonic crystal, we unveil a generalized bulk–edge correspondence intrinsically shaped by the relativistic nature of electromagnetic waves. This constraint imposes a strict frequency cutoff, a feature fundamentally absent in electronic topological insulators, which enables a new regime of frequency-controlled spatial localization near the cutoff. Furthermore,
we demonstrate that this generalized correspondence is polarization dependent: transverse electric (TE) and transverse magnetic (TM) edge modes exist in different parameter regimes and exhibit distinct dispersion relations, including different zero-dispersion points.
Our framework redefines the theoretical boundaries of topological photonics, unlocking
new opportunities for polarization-selective dispersion engineering and robust pulse propagation in topological photonic platforms.
\end{abstract}

\maketitle


\section{Introduction}\label{sec1}
Topological insulators, originally proposed in condensed-matter systems,
support localized edge states that are robust against disorder and
impurities due to topological protection~\cite{1,2,bernevig2013topological}.
These states appear at the interface between materials with distinct
topological invariants and persist even when the bulk remains insulating.
The emergence of these states is rooted in a fundamental and far-reaching
principle---the bulk--edge correspondence---which connects the
topology of the bulk bands to the existence of edge modes~\cite{Asboth2016}.
Over the years, the concept of topological edge modes has been extended
across diverse physical platforms, including acoustics~\cite{3},
mechanics~\cite{4,5}, ultracold atoms~\cite{6,7}, polariton condensates~\cite{8,9},
metamaterials~\cite{10}, and photonics~\cite{11,13,14}.
However, despite its broad applicability, the general validity of the bulk--edge correspondence remains an active area of research across a wide range of topological platforms, with a particularly strong interest in photonics.

Photonics has proven to be an especially fertile platform for realizing optical analogues of topological phases. Notable examples include systems featuring broken time-reversal symmetry that support unidirectional edge currents, structures with broken spatial inversion symmetry, such as valley Hall photonic, as well as various higher-order topological insulators, among others~\cite{11,12,13}. Within this landscape, the Su--Schrieffer--Heeger (SSH) model, noted for its conceptual simplicity and experimental versatility~\cite{15,PRB_SSH_1980,RMP_SSH_1988}, provides a minimal framework for studying topological phase transitions, designing tunable topological systems---including photonic crystals (PhCs)~\cite{30,16}---and exploring practical applications of robust edge propagation. Topological PhCs are realized in various geometries. For instance, in one-dimensional (1D) settings, most studies employ heterogeneous structures in which two topologically distinct PhCs are joined, giving rise to interface-localized topological modes~\cite{PRX2014_interface_modes,18,19,20}. These studies primarily focus on interface states rather than the canonical edge states defined in electronic topological insulators. In contrast, a systematic investigation of genuine photonic topological edge modes---particularly the conditions required for their existence and propagation, as well as their polarization-dependent and dispersive properties---remains largely unexplored. Addressing this gap is essential for clarifying the physical limits and practical applicability of topological protection in realistic photonic platforms, and for establishing design principles for topological devices tailored to specific applications. In communication-oriented architectures, strong modal confinement is required to preserve channel isolation and transmission fidelity. By contrast, sensing applications benefit from modes with sufficiently extended evanescent tails that enhance light–matter interactions by increasing overlap with the surrounding environment~\cite{Abood2025}. In both scenarios, device performance is governed by an interplay among topological protection, engineered dispersion, and modal localization~\cite{Kuipers_science2020}.

A proper understanding of the topological nature of these edge modes requires careful consideration of the bulk–edge correspondence.
While this principle is firmly established in electronic systems~\cite{Laughlin_PRB_1981,Thouless_PRL_1982,Halperin_PRB_1982,PRL1993_Hatsugai,PRB_2011_Mong_Shivamoggi}, its direct application to photonic platforms governed by Maxwell's equations demands deeper examination.
Although several recent works have studied this problem from a rigorous mathematical standpoint~\cite{denittis2017symmetryclassificationtopologicalphotonic,Lin_2022}, and physically motivated analyses have provided formal arguments for continuum Maxwell's equations in non-periodic media~\cite{PRX2019_Silveirinha} and finite 2D structures~\cite{qiu2025bulkedgecorrespondencefinitephotonic}, its status in periodic PhCs remains less clear. Typically, the bulk–edge correspondence in photonics is often invoked by analogy with electronic systems, despite the fundamental differences in the underlying dynamical equations and spacetime symmetries.
As a result, the precise conditions under which topological edge modes exist as physically realizable guided states remain not fully understood and warrant further investigation.

In this work, we address this issue by solving Maxwell's equations for an SSH-type PhC and demonstrating that the existence of topological edge modes is governed not only by conventional bulk topological invariants, but also by an additional guiding condition intrinsic to photonic systems. Specifically, edge modes must lie within the light-line region defined by the refractive indices of the constituent materials to remain localized. Although this requirement is well known in conventional waveguide theory, its role within topological photonic systems warrants special consideration. In conventional waveguides, modal confinement is primarily determined by the refractive-index contrast between the guiding core and the surrounding medium. In contrast, topological edge modes in photonic structures are shaped by the electromagnetic properties of the entire unit cell rather than by a single confining layer alone. As a result, the existence of localized edge states is restricted to a limited region of parameter space where both the topological and guiding conditions are simultaneously satisfied. This leads to a generalized form of bulk–edge correspondence for guided photonic modes. Within this framework, the non-zero value of conventional bulk topological invariants is no longer sufficient to guarantee the existence of localized edge states, but instead constitutes only a necessary condition whose physical realization depends on additional waveguiding constraints, in contrast to electronic systems, where such invariants directly ensure the presence of edge modes. This finding reveals a fundamental difference between photonic and electronic platforms rooted in their different underlying dynamical equations. Furthermore, we show that this generalized correspondence is inherently polarization dependent, as transverse electric (TE) and transverse magnetic (TM) modes exhibit distinct dispersion relations, cutoff conditions, and confinement regimes. Beyond their existence, our analysis reveals that the interplay between topology and wave propagation gives rise to a distinct guiding mechanism for edge modes, fundamentally different from conventional surface-state formation based on local refractive-index modifications. This enables transformative functionalities, including frequency-tunable spatial localization near the cutoff and unconventional dispersion regimes, such as zero-dispersion points, that transcend the traditional light-line constraints of standard waveguides. These results establish a predictive framework for engineering topological edge states in photonic platforms and open new routes for dispersion control and robust light manipulation in integrated photonic systems with potential applications in pulse shaping, soliton formation, and frequency conversion~\cite{Agrawal2013, Dudley2006, Ranka2000}.

\begin{figure}[tbp]
\centering \includegraphics[width=0.5\textwidth]{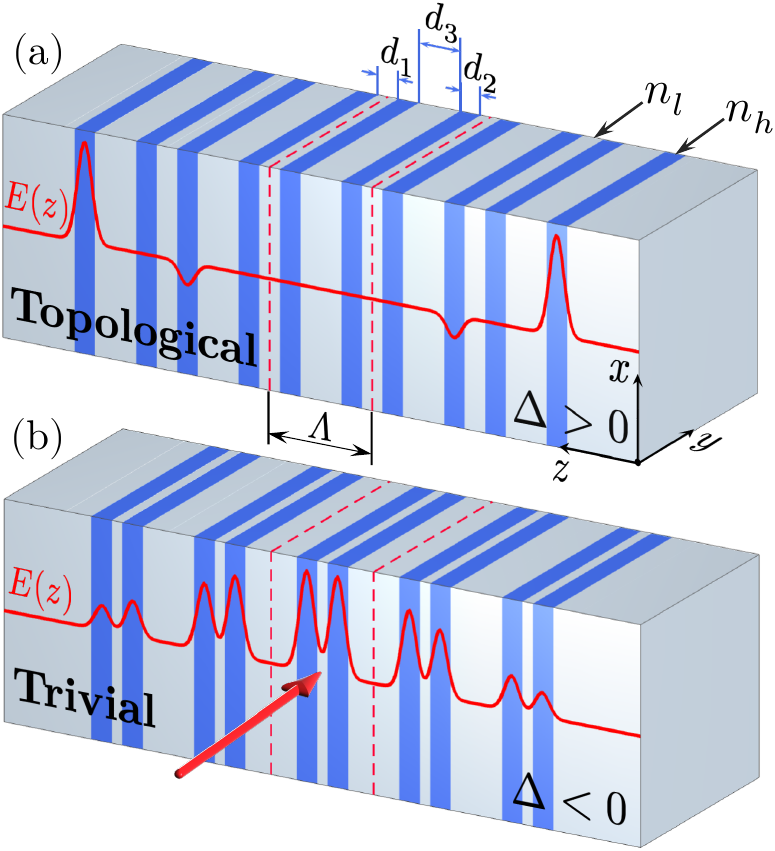}
\caption{Schematic of a PhC with planar waveguides of low ($n_{l}$)
and high ($n_{h}$) refractive indices, shown in topological (a) and
trivial (b) configurations. The refractive index is modulated along
$z$. The red dashed line marks the unit cell of period $\Lambda$,
with layer thicknesses $d_{1}$, $d_{2}$, and $d_{3}$.
In (a), the red curve depicts a topological edge mode; in (b), a bulk
mode. The large red arrow indicates the propagation direction.}
\label{structure} 
\end{figure}

\section{Theoretical model}
Our system is based on the 1D SSH model, with alternating hopping amplitudes mapped onto the layer thicknesses of the PhC. The distinction between trivial and nontrivial topological regimes is determined by these thicknesses, analogous to the electronic SSH model, where the inter- and intra-cell hopping coefficients define the system's topological character. Schematic representations of the two configurations are shown in Fig.~\ref{structure}, which also indicates the characteristic layer thicknesses $d_{1}$, $d_{2}$, and $d_{3}$ forming a five-layer unit cell ($d_{1}/2$, $d_{2}$, $d_{3}$, $d_{2}$, $d_{1}/2$) with period $\Lambda=d_{1}+2d_{2}+d_{3}$. To clarify the topological characterization and emphasize the analogy with the electronic model, we define the dimerization parameter as $\Delta=(d_{3}-d_{1})/4$: for $\Delta>0$ the system is in the topological phase [Fig.~\ref{structure}(a)], while for $\Delta\le 0$ it is in the trivial [Fig.~\ref{structure}(b)]. Because $\Delta$ depends only on $d_{1}$ and $d_{3}$, fixing $d_{2}$ and the period $\Lambda$ means that different values of $\Delta$ correspond to different combinations of $d_{1}$ and $d_{3}$, with $d_{1}$, $d_{3}>0$.

To capture the vectorial nature of light in structured media, we use the wave equations for the electric and magnetic fields in a linear, source-free, non-magnetic dielectric medium~\cite{31}: 
\begin{align}
\nabla\times\nabla\times\mathbf{E}(\mathbf{r}) & =\left(\frac{\omega}{c}\right)^{2}\varepsilon(\mathbf{r})\mathbf{E}(\mathbf{r})\,,\label{eq:wave_E}\\
\nabla\times\left(\frac{1}{\varepsilon(\mathbf{r})}\nabla\times\mathbf{H}(\mathbf{r})\right) & =\left(\frac{\omega}{c}\right)^{2}\mathbf{H}(\mathbf{r})\,,\label{eq:wave_H}
\end{align}
where $\omega$ is the angular frequency, $c$ is the speed of light, and $\varepsilon(\mathbf{r})$ is the spatially varying dielectric function. The SSH PhC is described by a 1D dielectric profile, $\varepsilon(\mathbf{r})\equiv\varepsilon(z)$, taking values $n_l^2$ in layers of thickness $d_1$ and $d_3$, and $n_h^2$ in layers of thickness $d_2$, as illustrated in Fig.~\ref{structure}. Throughout our analytical and numerical analyses, we use $n_l=1.60$ and $n_h=3.50$, corresponding to Al$_2$O$_3$ and GaAs, respectively~\cite{Suarez_2009}. Unless stated otherwise, the structural parameters $d_2$ and the period $\Lambda$ are fixed at $0.1$ and $0.5\,\mu\mathrm{m}$, respectively, while $d_1$ and $d_3$ are allowed to vary; their values can be directly obtained from the relations previously provided for $\Lambda$ and the dimerization parameter $\Delta$. We adopt the PhC convention for transverse electric (TE) and transverse magnetic (TM) modes classification: the TE (TM) mode corresponds to a nonzero magnetic (electric) field along the mirror symmetry axis (here, the $x$-axis), while the mirror plane is the $yz$-plane~\cite{31}. Notably, this classification is opposite to the standard waveguide convention~\cite{33}.

\begin{figure*}[t]
\centering \includegraphics[width=0.7\linewidth]{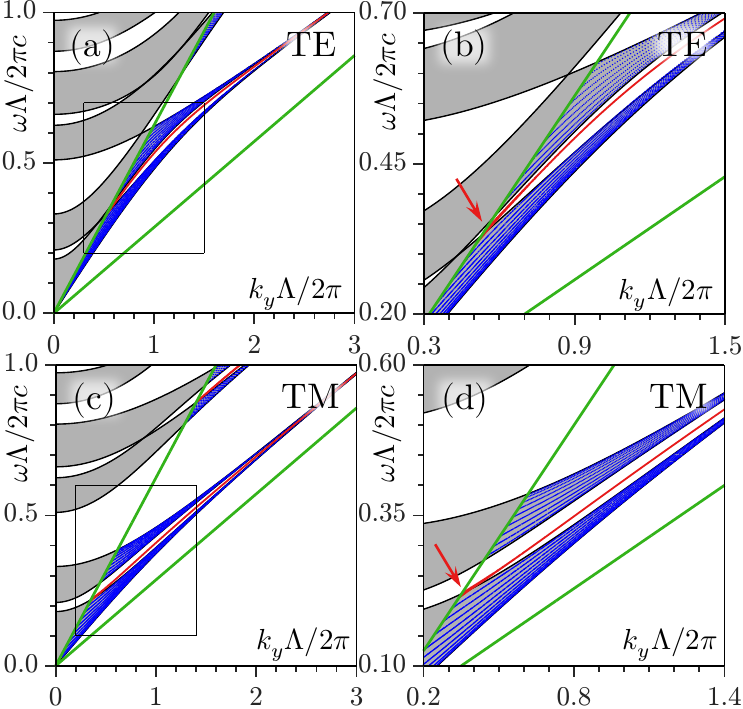}
\caption{Dependences of the normalized frequency $\omega\Lambda/2\pi c$ on
$k_{y}\Lambda/2\pi$ for an SSH PhC with $\Delta/\Lambda=0.05$. Panels
(a) and (b) correspond to TE polarization, and (c) and (d) to TM polarization. Panels (b) and (d) present magnified views of the regions marked by the black rectangles in (a) and (c), chosen to highlight the frequencies of the guided modes of the finite nine-period structure, including topological edge modes (red curves) and bulk modes (blue curves), confined between the light lines (green), defined by $\omega_{+}=ck_{y}/n_{\ell}$ and $\omega_{-}=ck_{y}/n_{h}$. Gray regions indicate the Bloch band structures of the periodic lattice. Red arrows in (b) and (d) indicate the transition points, corresponding to the cutoff values, at which the topological edge states disappear. Here, $d_{1}=d_{2}=0.1\,\mu\mathrm{m}$ and $d_{3}=0.2\,\mu\mathrm{m}$, resulting in a unit-cell period of $\Lambda=0.5\,\mu\mathrm{m}$.}
\label{band_structures}
\end{figure*}

\begin{figure*}[t]
\centering \includegraphics[width=0.7\textwidth]{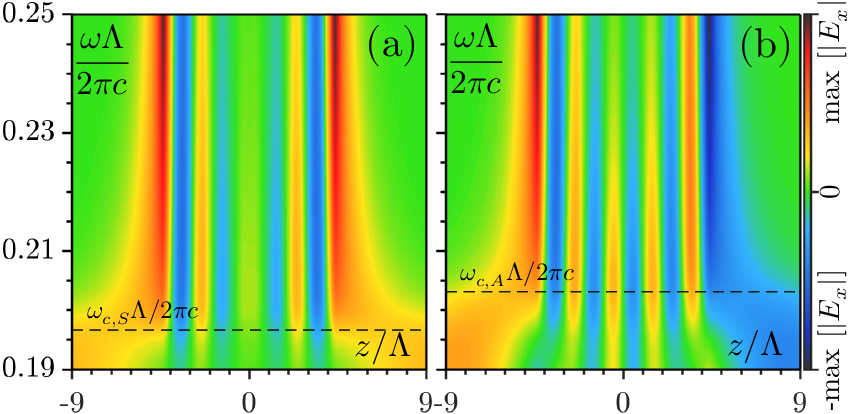}\caption{Spatial distribution of the electric field at different frequencies,
$E_{x}\left(z,\omega\right)$, for (a) the symmetric and (b) antisymmetric
edge TM modes. In this figure, we adopt a different set of layer thicknesses than the standard values used elsewhere in the manuscript, while all other parameters are the same as in the main text: $d_1=0.1 \Lambda$, $d_2=0.2 \Lambda$, $d_3=0.5 \Lambda$, with $\Lambda = 1\,\mu\mathrm{m}$, for a structure as in Fig.~\ref{band_structures} ($\Delta >0$) with 9 unit cells.
The localization--delocalization transition occurs
in the regions above and below the frequency cutoffs $\omega_{c,S}$
and $\omega_{c,A}$ represented by the dashed lines in each of the
panels, respectively. }
\label{fig:electric-field-w-vs-z}
\end{figure*}

\section{Existence criteria for topological photonic edge modes} 
In this section, we explicitly demonstrate that the bulk–edge correspondence conditions in photonic systems differ from those in electronic systems, and show that these differences directly affect the localization properties of edge states in PhCs.

\subsection{Bulk-edge correspondence for an SSH array of electronic channels}
We refer first to the classical formulation of the bulk–edge correspondence, as developed in the context of electronic systems (see, e.g., Asbóth et al. \cite{Asboth2016}, and the rigorous mathematical works of Graf and Porta \cite{Graf2013} for 2D systems and Shapiro \cite{Shapiro_2019} for 1D systems, including the SSH model). In these works, the bulk–edge correspondence establishes the equivalence between a bulk topological invariant or bulk index $\mathcal{I}_b$---defined for bulk (extended) eigenstates of the bulk Hamiltonian---and an edge index $\mathcal{I}_e$ that is calculated out of the edge Hamiltonian describing the finite system. The edge index is thus evaluated differently from the bulk index and counts the number of localized edge modes of the edge Hamiltonian. Importantly, the edge Hamiltonian is derived from the bulk Hamiltonian by introducing a single boundary to the latter, typically via Dirichlet conditions \cite{Graf2013,Shapiro_2019}. The standard bulk-edge correspondence establishes that, under proper assumptions, the bulk and edge indices are equal: $\mathcal{I}_b=\mathcal{I}_e$. For chiral-symmetric systems such as the SSH model, this correspondence is exact: the bulk invariant takes quantized values 0 or 1, which directly determine the number of exponentially localized edge states. Thus, in this framework, the topological non-triviality of the bulk Hamiltonian necessarily implies the existence of localized edge modes. Crucially, these results rely on the assumption that the system is described by a Hermitian Schrödinger operator in the tight-binding approximation with short-range (e.g., nearest-neighbor) hopping. 

In this situation, the bulk-edge correspondence also holds for an SSH array of electronic channels. In this case, the SSH modulation of the lattice potential occurs in the transverse direction, while the system remains unchanged along the axial direction, such that bulk and edge Hamiltonian eigenstates propagating axially also satisfy the bulk-edge correspondence. The reason is that, for an axially invariant Hamiltonian, say along the $y$ direction, all its eigenstates have a trivial dependence on $y$ and its conjugate variable $k_y$, namely $\psi_E(y,z)=\phi_E(z)e^{i k_y y}$, which does not modify the corresponding bulk and edge indices (see Sec.~5 of the SI). The bulk index remains unchanged because its calculation is independent of the trivial axial dependence of the propagating modes. The edge index is also unaltered since the number of edge-localized modes is independent of $k_y$, owing to the trivial dependence of the energy eigenstates on this quantity. In mathematical terms, this implies that $\mathcal{I}_b=\mathcal{I}_e$ is satisfied for all values of $k_y$. In all cases, both the Hamiltonian eigenvalues (energies) and the axial wave-vector component $k_y$ (the eigenvalue of the self-adjoint generator of axial translations) must be {\em real} numbers.

\subsection{Conditions for edge localization in an SSH photonic crystal waveguide}
Our analysis shows that, when considering propagating modes in an SSH-type photonic crystal waveguide, as shown in Fig.~\ref{structure}, the conventional formulation commonly used in electronic systems requires fundamental revision.
In particular, the relevant conditions for PhC can be stated as follows:

\begin{enumerate}
\global\long\def\labelenumi{\roman{enumi})}%
\item The bulk structure must reside in a topologically nontrivial phase. For the SSH PhC considered here, this requires $\Delta>0$, yielding a non-vanishing bulk index $\mathcal{I}_b^{ph}$ for the periodic (non-truncated) lattice~\cite{Asboth2016};
\item The localized edge modes of the truncated lattice must lie within the region bounded by the light lines, $\omega_{+}=ck_{y}/n_{l}$ and $\omega_{-}=ck_{y}/n_{h}$, where $k_y$ is the propagation constant and thus they must exhibit a cutoff in frequency.
\end{enumerate}

The first condition directly translates the bulk-edge correspondence principle from the SSH model to the photonic case. This guarantees a photonic non-zero bulk index, i.e., $I_b^{ph}\ne0$ when $\Delta>0$. The second condition requires the existence of an edge-localized mode with both a well-defined real frequency $\omega$ and a real axial wave-vector component $k_y$. In this sense, photonic edge modes satisfy criteria analogous to those of their electronic counterparts, namely edge localization accompanied by well-defined real energy and momentum. The number of localized photonic modes at a given edge satisfying this second condition can then be used as a basis for defining the photonic edge index $I_e^{ph}$ for the SSH PhC.

By solving the eigenvalue problem derived from Maxwell's equations~\eqref{eq:wave_E} and~\eqref{eq:wave_H}, we obtained the complete set of modes and their corresponding dispersion relations, $\omega_{n}(k_{y})$, for TE and TM polarizations for both the finite and infinite (bulk) structures (see details in Sec.~1 of the Supporting Information~(SI)). The gray shaded regions in Fig.~\ref{band_structures} indicate the bulk Bloch bands for TE [Fig.~\hyperref[band_structures]{\ref*{band_structures}(a)} and ~\hyperref[band_structures]{\ref*{band_structures}(b)}] and TM [Fig.~\hyperref[band_structures]{\ref*{band_structures}(c)} and ~\hyperref[band_structures]{\ref*{band_structures}(d)}] polarizations. 

Next, we proceeded to evaluate the photonic bulk index $\mathcal{I}_b^{ph}$ of the Bloch modes associated with the SSH PhC structure in Fig.\ref{structure}. We numerically verified that the topological invariant of these Bloch modes, expressed via the Zak phase~\cite{11,12,Asboth2016} and defining their bulk index $\mathcal{I}_b^{ph}$, is non-zero for $\Delta>0$ and has the form: $\mathcal{I}_b^{ph}= (-1)^{n+1}$, where $n$ is the band number. In contrast, the bulk index vanishes for $\Delta \le 0$: $\mathcal{I}_b^{ph}=0$ (see Sec.~2 of the SI).  Crucially for our discussion, this result is {\em independent} of the axial wave vector $k_y$.  In this way, the standard bulk--edge correspondence condition for the bulk index is satisfied: we indeed have a topologically non-trivial (infinite) bulk structure. However, fulfilling this topological condition alone does not {\em always} guarantee the existence of localized photonic edge modes in our case. This sets a new scenario to properly evaluate their number and, therefore, their corresponding edge index $\mathcal{I}_e^{ph}$.

To understand this essential characteristic of edge photonic modes and thereby determine their edge index, $\mathcal{I}_e^{ph}$, we carefully analyse the modal dispersion relations of the finite SSH PhC shown in Fig.~\ref{band_structures}. In all panels of this figure, the blue and red curves correspond to guided modes of the finite structure, constructed from $N=9$ periods, where blue represents the extended volume modes, and red indicates the topological edge modes.
Figs.~\hyperref[band_structures]{\ref*{band_structures}(a)} and~\hyperref[band_structures]{\ref*{band_structures}(c)} show that higher-order modes can appear in different frequency ranges for TE and TM polarizations, while Figs.~\hyperref[band_structures]{\ref*{band_structures}(b)} and~\hyperref[band_structures]{\ref*{band_structures}(d)} provide a zoom-in of the fundamental edge modes within the topological gaps for each polarization. As in electronic systems, the dispersion relations of the edge modes in the finite photonic structure appear within the topological gaps when $\Delta>0$ (see Sec.~3 of the SI). However, topological edge modes cease to exist as guided when $\omega_{\mathrm{edge}}\left(k_{y}\right)>\omega_{+}\left(k_{y}\right)$. The point where $\omega_{\mathrm{edge}}\left(k_{yc}\right)=\omega_{+}\left(k_{yc}\right)$ therefore defines both a lower cutoff for the propagation constant ($k_{y}>k_{yc}$ for guided modes) and also a corresponding lower cutoff frequency $\omega_{\mathrm{c}}=\omega_{\mathrm{edge}}\left(k_{yc}\right)$ (red arrows in Fig.~\hyperref[band_structures]{\ref*{band_structures}(b)} and~\hyperref[band_structures]{\ref*{band_structures}(d)}), below which the edge mode is no longer guided, as a result of violating condition~(ii) (see details in Sec.~4 of the SI). The failure to satisfy the guiding condition, $\omega_{-}\left(k_{y}\right)\le\omega_{\mathrm{edge}}\left(k_{y}\right)\le\omega_{+}\left(k_{y}\right)$, is therefore an extra delocalization mechanism for propagating modes. This loss of confinement is evident when examining the full set of edge mode field profiles near the cutoff frequencies, as shown in Fig.~\ref{fig:electric-field-w-vs-z}. In this parity-symmetric structure (invariant under the parity transformation $z\rightarrow-z$), the true edge eigenstates are the \emph{even} symmetric (S) and \emph{odd} antisymmetric (A) modes. At higher frequencies, both S and A modes are perfectly localized at the edge, analogous to propagating edge states in 2D SSH electronic systems (see Sec.~5 of the SI). For an increasingly large bulk width $l=N \Lambda$, the S and A modes become nearly degenerate and equivalent to an uncoupled pair of L and R modes fully localized at the edges (see details in Sec.~6 of the SI). In the case of a semi-infinite SSH PhC with only finite boundary---e.g., the left one---only one localized L edge propagating mode would survive (strictly speaking, a semi-infinite geometry is required for a rigorous definition of the edge index in order to isolate a single edge). This situation is precisely the one necessary to determine the edge index as in electronic systems \cite{Graf2013,Shapiro_2019}.

Thus, above the cutoff, edge localization occurs as in electronic systems, and an edge index $\mathcal{I}_e^{ph}=1$ can be associated with each edge of the SSH PhC. On the contrary, below the cutoff, i.e., for $k_y<k_{yc}$, or equivalently, for $\omega < \omega_{c,S}$ for symmetric modes and $\omega < \omega_{c,A}$ for antisymmetric modes, these states become leaky ($k_y$ becomes complex) and resonantly couple to radiative modes, losing their edge localization entirely. 
As a result, edge-localized modes with real $k_y$ and frequency $\omega$ are not found below the cutoff propagation constant $k_{yc}$. By analogy with electronic systems, this suggests that the edge index vanishes in this regime, i.e., $I_e^{ph}=0$ for $k_y<k_{yc}$. Importantly, TE and TM polarizations exhibit different cutoff frequencies. As a result, TE and TM polarizations can have different edge indices depending on $k_y$. In summary, the equality of bulk and edge indices characteristic of the conventional bulk-edge correspondence in electronic systems is not fully preserved in our SSH PhC waveguide and has to be properly adapted to the photonic case. When the bulk is topologically non-trivial, the bulk-edge correspondence $I_b^{ph}=I_e^{ph}$ remains valid only for $k_y>k_{yc}$, breaking down for smaller values than the critical propagation constant $k_{yc}$. As we will see below, this generalized bulk–edge correspondence in the SSH PhC has important physical consequences both for the degree of localization of edge modes and for pulse propagation.

\subsection{Localization-delocalization mechanism in an SSH photonic crystal waveguide}
In addition to the complete delocalization of edge modes and their conversion into leaky radiative modes below the cutoff frequency, the spatial distribution of the electric field at different frequencies, as illustrated in Fig.~\ref{fig:electric-field-w-vs-z}, unveils an additional mechanism by which the modes extend into the bulk of the SSH PhC. Unlike spreading into the outer region of the SSH PhC, this mechanism manifests as the formation of tails extending into the interior of the finite structure. At high frequencies, when $\omega\gg\omega_{c,S},\omega_{c,A}$, both S and A edge modes are strongly localized in the outermost SSH PhC waveguides, with negligible frequency separation between them in dispersion relations dependencies (see Sec.~6 of the SI). As the frequency approaches the cutoff region, the bulk tails of the edge modes expand, resulting in an increased frequency separation between the S and A modes near the transition region, $\omega \gtrsim \omega_{c,S}, \omega_{c,A}$. This separation originates from an effective $k_y$-dependent coupling $q$ between the left (L) and right (R) ends of the S and A edge modes of a finite SSH PhC, induced by the increased delocalization of the modes. In fact, the system in its topological phase can be described by a dimensionally reduced field theory in 1+1 dimensions involving weakly coupled L and R edge fields  (see Sec.~7 of the SI). The behavior of $q$ suggests that the localization--delocalization transition of edge modes in the SSH PhC can be interpreted as a crossover phase transition, arising from a $U_L(1) \otimes U_R(1)$ symmetry-breaking pattern associated with independent phase transformations $\theta_L$ and $\theta_R$ of the L and R fields. With $q\neq 0$, only ``vector'' transformations with $\theta_L=\theta_R$ remain symmetries, whereas ``axial'' transformations with $\theta_L \neq \theta_R$ do not. Thus, $q$ serves as an order parameter for ``axial'' symmetry breaking (see Sec.~7 of the SI for details).

Thus, we have shown that in photonic systems the bulk–edge correspondence is fundamentally different from that in electronic systems, and this difference leads to two distinct mechanisms of topological edge-mode delocalization. The origin of this distinction lies in the contrasting spacetime symmetries of the governing equations. In electronic systems, the dispersion relations of propagating states in the analogous 2D model (a SSH array of electronic channels in the $y$ direction) are quadratic in $k_{y}$, a direct consequence of the Galilean spacetime symmetry of the axial dependence on $y$ of the nonrelativistic Schrödinger equation (see Sec.~5 of the SI). This guarantees that, in the SSH electronic system, the standard bulk--edge correspondence $\mathcal{I}_b=\mathcal{I}_e$ is automatically satisfied for any propagating electronic edge mode and for all $k_{y}$, so that no additional criteria are required when $\Delta>0$. For an SSH PhC, the situation is inherently different: the relativistic spacetime symmetry of Maxwell's equations constrains guided modes to lie between the linear dispersion lines $\omega_{+}=ck_{y}/n_{l}$ and $\omega_{-}=ck_{y}/n_{h}$. This constraint, absent in the electronic case, is the reason behind the breakdown of the conventional bulk-edge correspondence in the SSH PhC.  These linear dispersion relations correspond to two effectively massless relativistic fields $\psi_{l,h}(y,t)$ propagating along the ${y}$-direction at speeds $c/n_l$ and $c/n_h$, showing Lorentzian invariance instead of the Galilean symmetry in the $y$ variable of all propagating states in the nonrelativistic Schrödinger equation of a SSH array of electronic channels. This essential distinction in the spacetime symmetries of the evolution equations in the two cases not only clarifies the conditions for the existence and transverse localization of photonic edge modes but also directly shapes their dispersion properties. Additionally, we have also alternatively checked these generalized edge localization conditions by solving Maxwell's equations in the time domain.

Finally, it is also relevant to place these mechanisms within the context of waveguide theory in PhCs. It is well-established that surface modes can be generated at the edges of one-dimensional and two-dimensional photonic crystals \cite{31}. However, these modes are generated by a local modification of the refractive index exclusively in the outer layers of the PhC, rather than in the bulk. In contrast to conventional waveguides---including those supporting surface modes, where modal propagation is primarily determined by the refractive-index contrast between the core and the surrounding medium---the existence of topological edge photonic modes depends on the global structure of the unit cell. In this case, the light lines are governed by the refractive indices of all materials composing the unit cell, rather than solely by that of the external medium. As a result, topological edge propagation cannot be attributed to a single material interface, but instead emerges from the collective properties of the periodic structure.

\begin{figure*}[t]
\centering \includegraphics[width=0.7\textwidth]{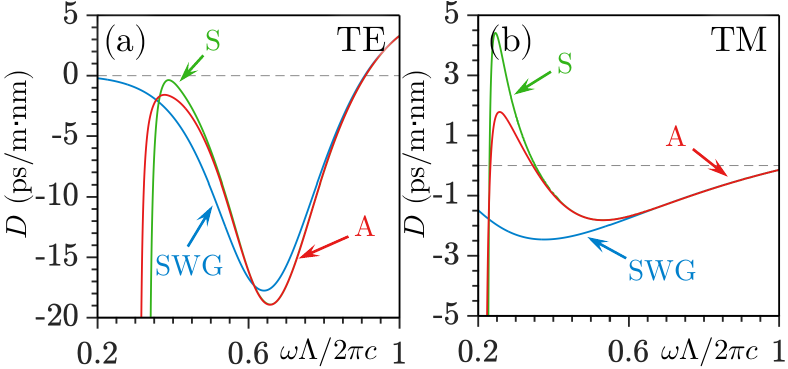}
\caption{Geometric dispersion coefficients $D$ as a function of frequency
for TE (a) and TM (b) polarizations for $\Delta/\Lambda=0.05$. The blue curves correspond to the fundamental mode of a single waveguide
(SWG), while the green and red curves represent the symmetric (S) and antisymmetric (A)
topological edge modes of the SSH PhC.}
\label{dispersion_coefficient_D} 
\end{figure*}

\section{Dispersion properties}
The consequences of this photonic bulk--edge correspondence become especially evident when examining light propagation. In particular, the constraints imposed by condition (ii) strongly influence the dispersion relations of topological edge modes in the nontrivial phase, enabling new opportunities for controlling pulse propagation. As shown in Fig.~\ref{band_structures}, the shape of the edge-mode dispersion relations (red curves) is affected both by the requirement that they remain within the bandgap---$\Delta>0$, condition (i)---and by the presence of cutoff frequencies [condition (ii), indicated by the red arrows]. For
high frequencies $\omega\gg\omega_{c,S},\omega_{c,A}$, where the effective coupling $q$ between the edge modes boundaries is negligible, the dispersion relations of the edge modes S and A almost overlap (quasi-degeneracy)
and become indistinguishable
from that of the fundamental single waveguide (SWG) mode, as can be seen in Fig.~\ref{dispersion_coefficient_D}. However, in the transition region $\omega\gtrsim\omega_{c,S},\omega_{c,A}$, the dispersion curves of the edge modes significantly approach those of the bulk modes of the lower
band, which remarkably modifies their slope and curvature and, consequently, their dispersion properties. 
The
standard parameter used in waveguide theory to quantify this effect
is the group velocity dispersion (GVD), $\beta_{2}\equiv d^{2}k_{y}/d\omega^{2}$
, or, alternatively, the dispersion coefficient $D=-2\pi c\beta_{2}/\lambda^{2}$
(see Sec.~8 of the SI for details). In Fig.~\ref{dispersion_coefficient_D}, we show the geometric dispersion coefficients $D$ for TE (a) and TM (b) modes. For both polarizations, we plot the dispersion $D$ for S and A modes. For comparison, we also include the $D$ curve of the fundamental mode of a single waveguide (SWG).
The effect in $D$ for the topological modes is apparent, particularly right above the transition region, when the curvature effects in the dispersion relation of topological edge modes due to the proximity to bulk modes are non-negligible.
Notably, we identify the positions of the zero-dispersion frequencies
for TM polarization, where the SSH PhC waveguide transitions between
normal ($\beta_{2}>0$) and anomalous dispersion ($\beta_{2}<0$)
regimes. In the anomalous dispersion regime, for example, the balance
between negative GVD and Kerr nonlinearity can support soliton formation~\cite{Agrawal2013}.
More broadly, engineered zero-dispersion points underpin supercontinuum
generation and enhanced nonlinear interactions in SSH PhC
systems~\cite{Ranka2000,Dudley2006}. 
Remarkably, this mechanism also applies to higher-order modes for both TE and TM polarizations (see the dispersion relation of a higher-order topological mode in the upper bandgap in Fig.~\hyperref[band_structures]{\ref*{band_structures}(c)} for TM polarization).

\section{Conclusions}
We have shown that in an SSH-inspired photonic crystal, a nontrivial bulk index alone does not guarantee the existence of topological edge modes. In contrast to the electronic SSH model, where edge states are ensured whenever the topological invariant is non-zero, the photonic case requires an additional guiding criterion: edge modes must lie within the region bounded by the light lines determined by the lowest and highest refractive indices of the structure, and outside this region, guided edge modes cannot exist.

As edge modes approach these light-line boundaries, they exhibit strongly modified confinement and dispersion, including pronounced polarization-dependent effects. In particular, TM modes can display zero-dispersion points near cutoff, while analogous regimes for TE modes can be engineered through structural design. This introduces an additional degree of control (absent in electronic systems) where polarization, geometry, and cutoff physics jointly shape the behavior of topological states. These effects enable new possibilities for dispersion engineering in topological photonics, including robust pulse shaping, soliton-supporting propagation, and enhanced nonlinear frequency conversion~\cite{Smirnova2020}.

Crucially, the emergence of these modes is fundamentally different from conventional photonic-crystal surface states, which rely on local refractive-index modifications at the boundary~\cite{31}. Here, all unit cells remain identical, and localization arises solely from a nontrivial bulk topology induced by dimerization, with the modes occupying the bandgap opened by the dimerization itself.

Beyond waveguide configurations, these results naturally extend to resonant photonic systems. In particular, SSH-type edge modes can be harnessed in Fabry–P\'{e}rot cavities and microresonators, where topology-assisted confinement offers a route to simultaneously tailoring mode volume and dispersion. This suggests new strategies for integrated laser architectures, enhanced light–matter coupling, and engineered nonlinear dynamics, including supercontinuum generation and frequency-comb formation. More broadly, these unique properties admit an interpretation in terms of symmetry patterns characteristic of $1+1$ dimensional field theories, establishing connections with well-known low-dimensional models in particle and condensed matter physics~\cite{Tong2018}.

\section*{Supporting Information}

\noindent The Supporting Information includes: 1. Wave Equations for $\mathbf{H}$ and $\mathbf{E}$. 2. Chern number calculation of the bulk structure. 3. Frequency dependency on dimerization parameter: tuning knob for system's topology. 4. Cutoff frequencies for edge modes in topological photonic crystals. 5. Propagating states in a 1D SSH potential in a 2+1 Schrödinger equation. 6. Localization-delocalization transition for topological edge modes. 7. Effective 1+1 Hamiltonian model for L and R edge modes. 8. Optical dispersion parameters. This also includes Refs.~[1, 4, 13, 16, 37--39, 46--51].

\section{Funding} 
This work received funding through projects no. PID2020‐120484RB‐I00, financed by the Spanish MCIN/AEI. This study also formed part of the Advanced Materials and Quantum Communication programs and was supported by MCIN with funding from European Union Next Generation EU (PRTR‐C17.I1) within the project COMCUANTICA/009. Besides, this work was supported by the Generalitat Valenciana PROMETEO/2021/082, and the European Union through the Program Fondo Social Europeo Plus 2021–2027 (FSE+) of the Valencian Community (Generalitat Valenciana CIAPOS/2023/329). Project developed within the framework of the General Collaboration Agreement between the University of Valencia and Banco Santander, S.A., for the Santander Postdoctoral Research Scholarships.


\nocite{rylander2012computational,JinJianMing2014,11,12,Asboth2016,Yeh1988,1,Altland2010a,Chalyi2025,Agrawal2013,Ranka2000,Dudley2006}











\bibliography{refs-arxiv}

\end{document}